\documentclass[12pt,preprint]{aastex}

\begin{document}

\title{Kinematics of X-ray Emitting Components in Cassiopeia A}

\author{Tracey DeLaney\altaffilmark{1}, Lawrence Rudnick\altaffilmark{1}, 
Robert A. Fesen\altaffilmark{2}, T. W. Jones\altaffilmark{1}, Robert 
Petre\altaffilmark{3}, and Jon A. Morse\altaffilmark{4}}

\altaffiltext{1}{Department of Astronomy, University of Minnesota, 116 Church 
Street SE, Minneapolis, MN  55455; tdelaney@astro.umn.edu, 
larry@astro.umn.edu, twj@astro.umn.edu}
\altaffiltext{2}{Department of Physics \& Astronomy, Dartmouth College, 
Hanover, NH  03755; fesen@snr.dartmouth.edu}
\altaffiltext{3}{NASA Goddard Space Flight Center, Greenbelt, MD  20771; 
petre@lheavx.gsfc.nasa.gov}
\altaffiltext{4}{Department of Physics \& Astronomy, Arizona State 
University, Box 871504, Tempe, AZ  85287-1504; Jon.Morse@asu.edu}

\begin{abstract}

We present high-resolution X-ray proper motion measurements of Cassiopeia A 
using \emph{Chandra X-ray Observatory} observations from 2000 and 2002.  We 
separate the emission into four spectrally distinct classes: 
Si-dominated, Fe-dominated, low-energy-enhanced, and continuum-dominated.  
These classes also represent distinct spatial and kinematic components.  The 
Si- and Fe-dominated classes are ejecta and have a mean expansion rate of 
0.2\% yr$^{-1}$.  This is the same as for the forward shock filaments but 
less than the 0.3\% yr$^{-1}$ characteristic of optical ejecta.  The 
low-energy-enhanced spectral class possibly illuminates a clumpy circumstellar 
component and has a mean expansion rate of 0.05\% yr$^{-1}$.  The 
continuum-dominated emission likely represents the forward shock and consists 
of diffuse circumstellar material which is seen as a circular ring around the 
periphery of the remnant as well as projected across the center.

\end{abstract}


\section{Introduction}

Cassiopeia A (Cas A) is the youngest known ($\sim$ 330 yr, Thorstensen, 
Fesen, \& van den Bergh 2001) Galactic supernova remnant (SNR).  It is about 
3.4 kpc \citep{rhf95} away and was most likely the result of either a type 
Ib or IIn supernova explosion \citep{co03}.  As one of the brightest X-ray 
and radio sources in the sky, Cas A is one of the best targets for studying 
SNR evolution.  At radio, X-ray, and optical wavelengths, Cas A is dominated 
by a 3$\farcm$5 diameter bright ring of emission with a fainter 5$\arcmin$ 
diameter plateau of radio and X-ray emission.  The X-ray emission is 
dominated by a thermal spectrum rich in emission lines from highly ionized 
atoms.  The radio emission is consistent with synchrotron radiation from 
relativistic electrons accelerated at shocks.  The optical emission 
originates from a complex system of chemically enriched knots.  

The kinematics of small and large scale radio features of Cas A are well 
studied.  The small-scale radio knots are significantly decelerated relative 
to the X-ray and optical emission.  They are not in homologous expansion and 
there are even inward moving knots.  The expansion rates of the small-scale 
features of the bright ring vary by over a factor of two as a function of 
azimuth with a mean rate of 0.11\% yr$^{-1}$ \citep{ar95,krg98,tuf86}.  
The bulk expansion of Cas A has been measured in several ways by 
\citet{ar95}, \citet{ag99}, and \citet{dr03} with reported expansion rates 
of $\approx$0.11\% yr$^{-1}$, $\approx$0.22\% yr$^{-1}$, and 
$\approx$0.07\% yr$^{-1}$, respectively.  For this discussion, we adopt the 
\citet{dr03} value of 0.07\% yr$^{-1}$ because it is the least contaminated 
by brightness changes and azimuthal asymmetries.

The optical components of Cas A consist of slow-moving ($\lesssim$ 500 
km s$^{-1}$) quasi-stationary flocculi (QSFs) and fast moving knots (FMKs, 
4000$-$6000 km s$^{-1}$, $\approx$0.3\% yr$^{-1}$) \citep{kv76,vk85}.  The 
QSFs are thought to be shocked circumstellar material (CSM) from the 
presupernova wind of the progenitor while most of the FMKs are located on the 
bright ring and represent emission from ejecta that have interacted with the 
reverse shock.  There are also outlying knots of enriched composition with 
velocities from 8000$-$15,000 km s$^{-1}$ ($\approx$0.3\% yr$^{-1}$) 
\citep{fes01}.  

Previous X-ray proper motion measurements, conducted with \emph{Einstein} 
and \emph{ROSAT} at a resolution of 5\arcsec, showed that the mean 
expansion rate was 0.2\% yr$^{-1}$ \citep{krg98,vbk98}.  This established a 
paradoxical difference in expansion rates for the cospatial bright ring 
measured in optical, X-ray, and radio of approximately 3:2:1, respectively 
(Thorstensen, Fesen, \& van den Bergh 2001 and references therein).  The 
factor of two discrepancy between the X-ray and radio expansion is a common 
theme in young SNRs and has been found in both Kepler's SNR 
\citep{hug99,dsa88} and Tycho's SNR \citep{hug00}.  

With the advent of the \emph{Chandra X-ray Observatory}, we are now able to 
measure proper motions with unprecedented spatial resolution.  The results 
for the forward shock filaments around the outer edge of the remnant showed a 
mean expansion rate of 0.21\% yr$^{-1}$ \citep{dr03}, the same as the bright 
ring.  In this paper we extend our proper motion analysis to the rest of the 
SNR, for four spectrally distinct components.  One of these components 
represents a new class of X-ray emission while three components have been 
previously identified \citep{hrb00}.  We interpret the proper motions based 
on these spectral components and propose a solution to the expansion rate 
paradox.

\section{Observations, Analysis, and Results}

Cas A was observed for 50 ks with the ACIS-S3 chip onboard
\emph{Chandra} on 2000 January 30-31 and 2002 February 6.  For details
of the observing parameters and data reduction see \citet{hhp00},
\citet{gkr01}, and \citet{dr03}.  To measure proper motions, we
registered the 2000 and 2002 broadband (0.3$-$10 keV) X-ray images by
aligning them on the point source as described in \citet{dr03}.  The
uncertainty of the registration to the point source is 0$\farcs$015 based on 
Monte Carlo simulations which translates to a proper motion error of 0.005\%
yr$^{-1}$ at the mean radius of the exterior forward shock filaments
(150\arcsec).  The possible 0$\farcs$02 yr$^{-1}$ \citep{tfv01} motion
of the point source would result in a maximum proper motion error of
0.013\% yr$^{-1}$ at 150$\arcsec$.  We constrained rotation between the two 
images by requiring the bright knots in the northeast jet to have no 
average rotation.  The jet knots are the least decelerated features in the 
remnant and so should be the least affected by CSM interactions.  The 
uncertainty in rotation alignment is $<0\fdg05$.  The proper motion 
measurements for 261 knots and filaments were then made by minimizing 
$\chi^2$ for a $\approx$7$\arcsec$ region around each feature, using the 
difference maps between the 2000 image and a set of 2002 images, shifted over 
a range of $\pm 2.5\arcsec$ in both RA and DEC. \cite{dr03} contains a
detailed description of the one-dimensional analog of this procedure.
We assigned an error of 0\farcs07 yr$^{-1}$ to all 261 features, which 
corresponds to the top 20\% of the Monte Carlo errors derived in \citet{dr03} 
where the mean error was 0\farcs04 yr$^{-1}$.  These Monte Carlo simulations 
include problems that would arise from features of varying sizes and the 
different point spread functions across the remnant, such as affects the 261 
features measured here.  The positions and proper motions for the 261 
features are shown in Table \ref{tbl1}.  We determined the radial component 
of the motion for each knot or filament and calculated the radial expansion 
rate ($v_r$) using the distance from the optical expansion center at 
$23^{\mathrm{h}}23^{\mathrm{m}}27\fs77$, 
$+58\degr48\arcmin49\farcs4$ (J2000) \citep{tfv01}.  

The 261 knots and filaments were spectrally classified using both extracted 
spectra and spectral tomography, as discussed below.  Based on the extracted 
spectra, four major classes were identified:  Si-dominated, 
Fe-dominated, low energy (LowE) enhanced, and continuum-dominated.  
The Si-, Fe-, and continuum-dominated classes were previously identified by 
\citet{hrb00}.  Representative spectra are shown in Figure \ref{spectra}.  
The Si-dominated emission is marked by very strong line emission from Si and 
S with contributions from O, Ca, Ar, and Mg.  The Fe-dominated spectrum shows 
strong FeL and FeK emission and can in some cases be modeled with nearly pure 
Fe emission \citep{hl03,lh03}.  The LowE-enhanced emission shows a relative 
increase in emission at low energies ($\approx$ 0.8$-$1.6 keV) compared to 
the global spectrum.  The continuum-dominated emission is marked by very 
little or no line emission and an increase in 4$-$6 keV emission.  Only 50\% 
of the knots and filaments can be cleanly classified using their spectra 
alone.  The remaining features combine the characteristics of two or more of 
the four basic classes identified above.

To separate overlapping spectral structures and to spatially visualize the 
spectral components, we used spectral tomography.  The spectral tomography 
technique has been used to analyze the spectra of radio galaxies \citep{kr97} 
and supernova remnants \citep{dkr02}.  It involves taking differences between 
images from two different energies with a scale factor chosen to accentuate 
features of interest.  In Figure \ref{tomos} we show four tomography images 
each representing the dominant spectral type indicated by the spectra in 
Figure \ref{spectra}.  To make the tomography images, we first constructed 
emission line images for Si (1.72$-$1.96 keV which includes Al and possibly 
weak Mg), FeL+Ne (0.9$-$1.15 keV which includes Ni), and FeL+Mg (1.35$-$1.55 
keV) and an image of the high energy continuum (HEC) region from 4$-$6 keV.  
These energy ranges were chosen to exploit the differences between the 
spectral classes.  The images were convolved to 3$\arcsec$ resolution and 
there was no background subtraction or correction applied for differential 
absorption across the remnant.  The Si-dominated image was constructed by 
scaling and subtracting the HEC image from the Si image.  Regions that are 
bright on the Si-dominated image represent ejecta whose emission is dominated 
by strong emission lines of the Si-group (Si, S, Ar, Ca) which all scale with 
the Si line \citep{hhp00}.  Regions that are dark on the Si-dominated image 
represent features where the line emission is not dominant with respect to 
the continuum.  The scale factor was chosen to separate the bimodal 
distribution of ratios in the log(Si/HEC) image.  The appearance of the image 
in Figure \ref{tomos} is only weakly dependent on the value of the scale 
factor.  Because only the HEC image was subtracted to create the Si-dominated 
image, there are low levels of contamination from the Fe-dominated and 
LowE-enhanced classes.  Subtracting out these two components does not 
significantly alter the Si-dominated image because the relative strength of 
the Si emission is very strong compared to the other classes.  

The continuum-dominated image was made in the same way as the Si-dominated 
image, but with a slightly different scaling factor to better show the 
continuum-dominated component.  The brightness on the image was reversed so 
the continuum-dominated emission appears positive while the Si-dominated 
emission appears negative.  The Fe-dominated image was constructed from 
scaling and subtracting the FeL+Mg image from the FeL+Ne image.  The bright 
areas represent regions where the ratio of FeL+Ne to FeL+Mg is greater than 
one, which is unique to the Fe-dominated ejecta class.  The LowE-enhanced 
image was made by first scaling and subtracting the Si image from the FeL+Ne 
image.  Because the resulting image also contained significant Fe-dominated 
ejecta emission, a second scaling and subtraction was applied using the 
positive brightness areas on the Fe-dominated image to eliminate regions 
with strong Fe ejecta emission.  The resulting image only shows those regions 
with spectra similar to the LowE-enhanced spectrum in Figure \ref{spectra}.  

The Si-dominated and Fe-dominated tomography images basically reproduce the 
spatial distributions of Si and Fe as shown by the equivalent width images 
in \citet{hhp00}, with the exception that the western region does not show as 
much Si and Fe emission as is actually present due to the increased 
absorption of low energy X-ray emission in that region \citep{kra96,wbv02}.  
The continuum-dominated image shows the same forward shock filaments around 
the outer edge of the remnant as described by \citet{gkr01}.  There is also 
continuum-dominated filamentary emission interior to the bright ring of the 
remnant which \citet{hrb00} speculated results from X-ray synchrotron emission 
associated with the forward shock.  The continuum-dominated image matches the 
8-15 keV \emph{XMM} continuum image \citep{bwv01}, in the west particularly, 
indicating that this emission may indeed have a strong nonthermal component.  

The LowE-enhanced image represents emission that has not previously been 
explored.  Some of the LowE-enhanced emission is associated with optical QSF 
emission such as the features in the interior of the remnant, the bright 
northeast region at the outer edge of the bright ring, and the southern arc 
\citep{lmu95}.  Some of the LowE-enhanced emission has no QSF counterparts 
such as the emission to the southeast.  This may be because the low energy 
spectral region contains emission lines of O, Fe, Ne, and Mg, making it 
possible that some of the emission in the LowE-enhanced image is due to more 
than one spectral component.

The power of the tomography technique is in its ability to isolate spatially 
overlapping spectral components.  Using spectra alone, one may misidentify 
features as intermediate spectral types when they are simply a combination of 
unrelated overlapping structures.  The fact that the tomography images 
isolate X-ray structures that match emission features in other wavebands 
proves the efficacy of the technique.  This will be explored in more detail 
in a future paper (DeLaney et al. in preparation).

\begin{figure}
\epsscale{0.5}
\plotone{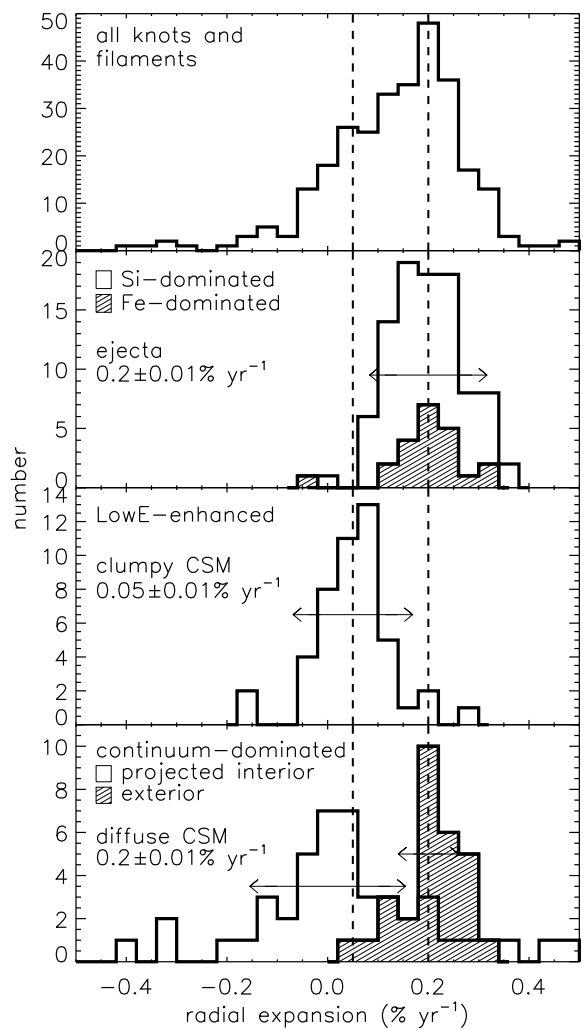}
\caption{Histogram of expansion rates in Cas A.  Arrows represent the average 
measurement error for each population.  Dashed lines represent the mean 
expansion values for the Si-dominated and LowE-enhanced populations.  There 
are also two continuum-dominated knots with expansion rates less than 
$-$0.4\% yr$^{-1}$.  The ``exterior'' continuum-dominated filaments are from 
\citet{dr03} and have been included here for completeness.  Free expansion 
corresponds to 0.3\% yr$^{-1}$.
\label{hist}}
\end{figure}

Using both extracted spectra and spectral tomography, 80\% of the 261 knots 
were classified into one of the four spectral classes.  Histograms showing 
the radial expansion rates broken down by spectral class are shown in Figure 
\ref{hist}.  Also included for completeness are the expansion rates from the 
29 forward shock filaments presented in \citet{dr03}.  The expansion rate 
distribution of the knots and filaments is very broad; however, when broken 
down by spectral class, it is clear that each species has its own 
characteristic motion.  The expansion rate distribution of each class holds 
for both the tomography and spectrum identified sets independently -- proving 
the robustness of the tomography identifications.  The Si- and Fe-dominated 
ejecta knots have the fastest expansion rates with a mean value of 
0.2$\pm$0.01\% yr$^{-1}$.  The expansion rate, expressed in \% yr$^{-1}$, is 
independent of projection effects.  

To compare to Doppler measurements, we can project the expansion rates onto 
the plane of the sky.  The Si-dominated knots are at an average radius of 
95$\arcsec$ which converts to a transverse velocity of 3100 km s$^{-1}$ for a 
distance of 3.4 kpc to Cas A.  The Fe-dominated knots lie further out on 
average at a radius of 120$\arcsec$ yielding an average transverse velocity 
of 3900 km s$^{-1}$ for the same expansion rate.  The Doppler measurements 
from \emph{XMM} show that the Fe emission is at a larger radius on average 
than the Si emission and that the average line-of-sight velocities are 
smaller than the average transverse velocities at approximately 1500 
km s$^{-1}$ for the Fe and 1000 km s$^{-1}$ for the Si \citep{wbv02}.  The 
Doppler velocities measured for Si with \emph{Chandra} are slightly faster 
than those measured with \emph{XMM} at 2000-3000 km s$^{-1}$ \citep{hsp01}.  
The difference between the Doppler measurements for Si could be due to the 
way in which the zero point in velocity was determined or possibly to the 
different spectral and spatial resolutions of \emph{XMM} and \emph{Chandra}.  
The mismatch of Doppler velocities and transverse velocities is not 
surprising.  The average transverse and line-of-sight velocities would only 
match if the ejecta were uniformly distributed and thus observed at all 
projected radii such that, on average, projection effects would contribute 
equally in each case.  For Cas A, the ejecta do not appear to be uniformly 
distributed \citep{wbv02}.  

The LowE-enhanced knots are moving quite slowly with a mean expansion rate of 
0.05$\pm$0.01\% yr$^{-1}$.  The Si- and Fe-dominated knot distribution and 
the LowE-enhanced knot distribution are narrow with respect to the separation 
between them which is consistent with virtually non-overlapping populations.  
The continuum-dominated knots and filaments are split into ``exterior'' and 
``projected interior'' where the exterior features are the forward shock 
filaments around the periphery of the remnant.  The exterior features have a 
mean expansion rate of 0.2$\pm$0.01\% yr$^{-1}$ while the expansion rates of 
the projected interior population varies significantly and includes apparent 
inward motions.  Most of the inward motions are concentrated to the south and 
west between 170$\degr$ and 300$\degr$ azimuth.  Radio proper motion 
measurements also show inward moving features in this region \citep{ar95}.

\section{Discussion}

As was found previously with the \emph{Einstein} and \emph{ROSAT} proper 
motion measurements \citep{krg98,vbk98}, the X-ray ejecta are moving slower, 
as a class, relative to the cospatial \emph{population} of optical ejecta and 
moving faster, as a class, than the cospatial \emph{population} of radio 
knots (On a case-by-case basis, the small number of matched X-ray/optical and 
X-ray/radio knots show the same motions (DeLaney et al. in preparation)).  
The X-ray/optical population difference is thought to be due to the density 
difference between the two populations \citep{ajr94,hsp01,fes01,dr03}.  High 
density ejecta experience very little deceleration, and thus very little 
disruption, allowing the ejecta to cool rather quickly and emit at optical 
wavelengths.  Low density ejecta experience more deceleration, and are heated 
to higher temperatures.  These ejecta emit at X-ray wavelengths and may even 
be completely disrupted before they can cool enough to emit optically.  The 
radio emission arises from synchrotron radiation, which requires 
amplification of magnetic fields to be strong enough for the emission to be 
seen. This amplification results from the turbulence associated with ejecta 
deceleration. If significant deceleration is needed for amplification, radio 
emitting knots would be decelerated the most. 

Of particular interest is the fact that the Si- and Fe-dominated ejecta have 
experienced the same fractional deceleration in their expansion rate from the 
free expansion rate of 0.3\% yr$^{-1}$ to 0.2\% yr$^{-1}$.  The Fe-rich 
emission (which results from explosive silicon burning near the core of the 
progenitor) is on average exterior to the Si-group emission (which results 
from explosive oxygen burning farther out from the star's center).  This 
spatial overturning of material could be due to several scenarios such as:  
1.  The Si and Fe ejecta may have had different initial velocities imparted 
from the SN explosion such that the Fe layer overtook the Si layer.  2.  The 
Fe ejecta may have experienced less deceleration at the reverse shock than 
the Si ejecta (perhaps due to density differences between the two types of 
ejecta).  3.  The overturning may have occurred in the star prior to or during 
the SN explosion resulting in a spatial separation between the Fe and Si even 
in the event of the same initial explosion velocities.  These scenarios would 
not necessarily predict that the Si and Fe would have the same expansion 
rate, as observed.  Furthermore, the X-ray ejecta have the same expansion 
rate as the forward shock filaments (i.e. homologous expansion on average), 
perhaps implying that the two populations are part of a coupled dynamical 
system in which the pressure in the X-ray ejecta is determined by the forward 
shock speed; although this does not explain why the optical ejecta would not 
be affected.  Even if initially the different types of ejecta were not 
coupled to eachother or to the forward shock speed, after interaction with 
the reverse shock and secondary shocks, the ejecta may now be coupled and 
expanding with the forward shock \citep{lh03}.

At the other end of the expansion rate spectrum are the LowE-enhanced knots 
and filaments.  Spatially, many of these features match very well with 
optical QSF emission, particularly the arc to the south, the emission on the 
bright ring to the northeast, and much of the interior emission 
(DeLaney et al., in preparation).  On average, QSFs show little or no motion 
\citep{vk85} while the mean expansion rate of the LowE-enhanced knots is 
0.05$\pm$0.01\% yr$^{-1}$.  Based on the spatial and dynamical similarities, 
we propose that the LowE-enhanced X-ray emission, like the optical QSF 
emission, represents the clumpy component of the CSM that was deposited by 
the wind from the progenitor.  The slight difference in expansion rates 
between the optical QSF population and the X-ray LowE-enhanced population 
might be due to a pre-existing density distribution in the progenitor wind 
\citep{co03}.  On average, less dense X-ray clumps may be accelerated more by 
the forward shock than more dense optical clumps.  The low-energy enhancement 
in the X-ray spectrum relative to the global X-ray spectrum (i.e. a 
``softer'' X-ray spectrum) is expected as dense clumps are overrun by the 
forward shock resulting in slow moving transmitted shocks through the clumps 
\citep{co03} which will produce an X-ray spectrum with a ``cooler'' 
characteristic temperature compared to other X-ray components 
\citep{hc86,pbg03}.

\begin{figure}
\epsscale{0.5}
\plotone{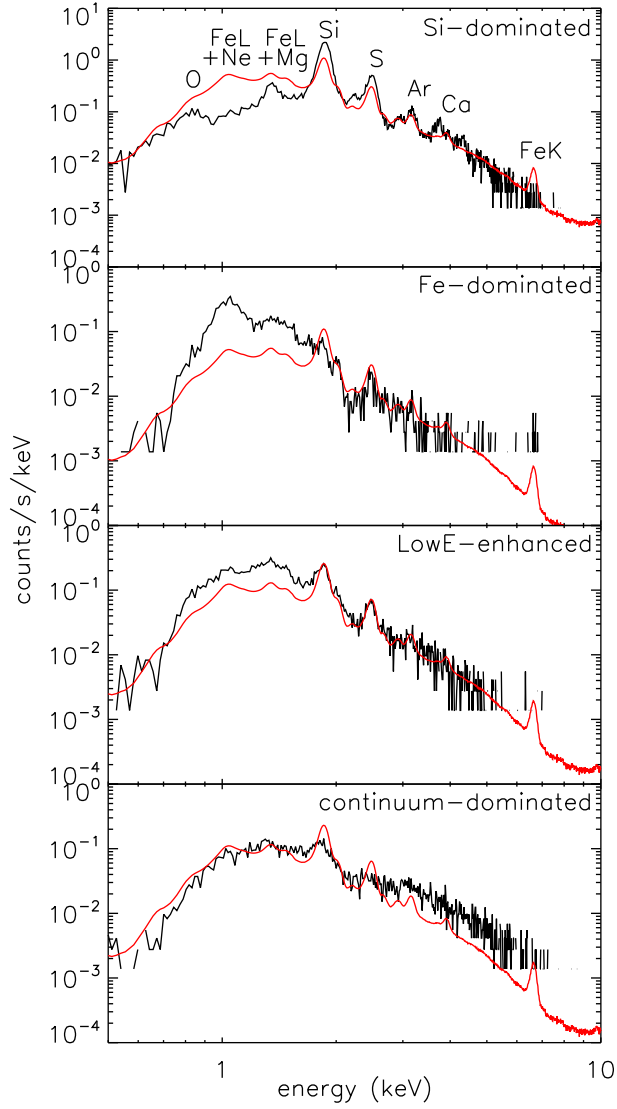}
\caption{\emph{Chandra} ACIS-S spectra of Cas A showing the global spectrum 
(in red) and typical spectra for the four spectral classes identified in the 
text.  The Si-dominated emission is marked by very strong Si group emission.  
The Fe-dominated spectrum shows strong FeL and FeK emission.  The 
LowE-enhanced emission shows a relative increase in emission at low energies 
compared to the global spectrum.  The continuum-dominated emission is marked 
by very little or no line emission and an increase in 4$-$6 keV emission 
relative to the global spectrum.
\label{spectra}}
\end{figure}

\begin{figure}
\epsscale{1}
\plotone{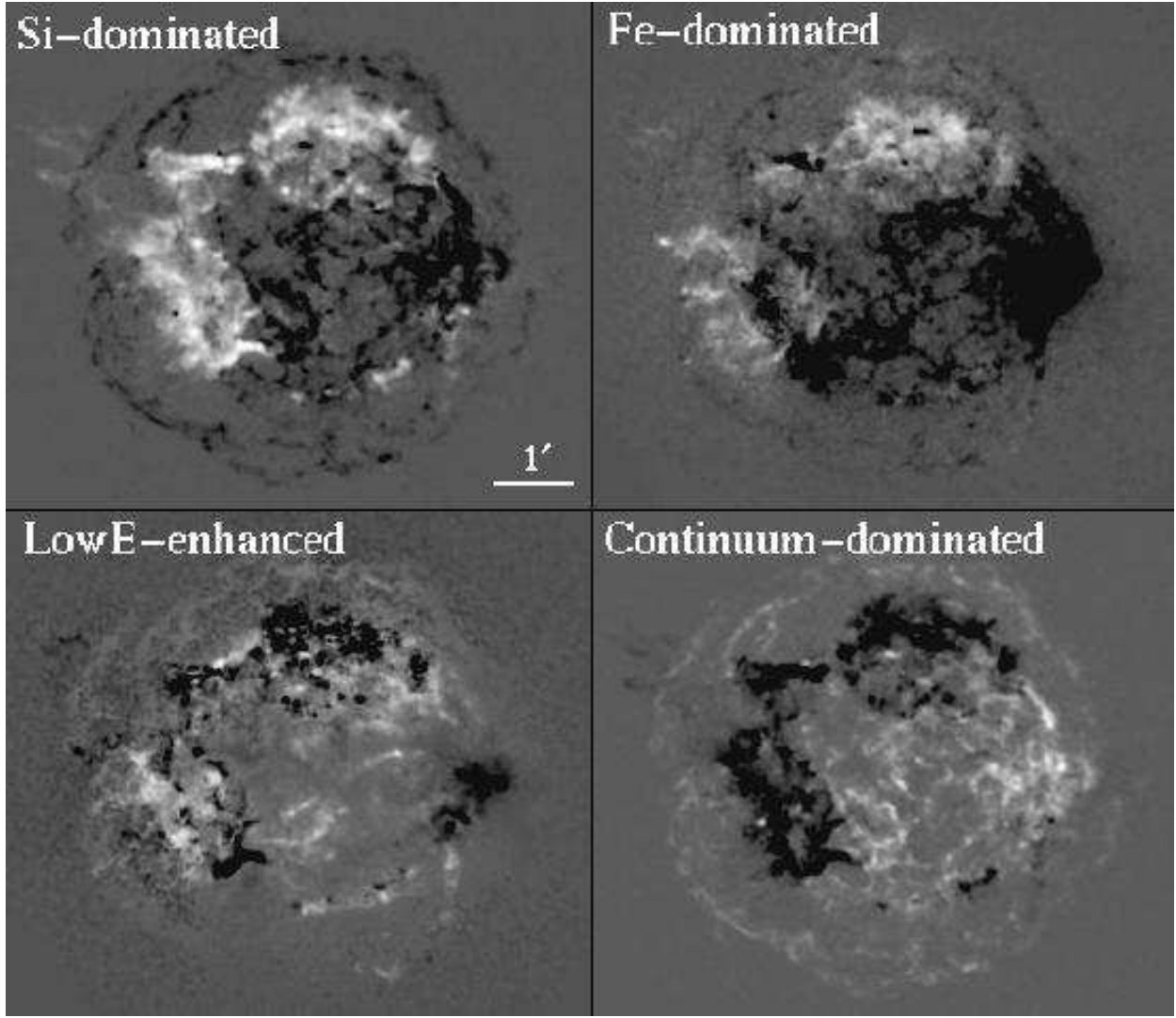}
\caption{Representative tomography images used to spatially identify 
emission from the four classes whose spectra are shown in Figure 
\ref{spectra}.  The Si- and Fe- dominated images represent ejecta emission.  
Note that, because of absorption, the ejecta emission on the western side of 
the remnant is underrepresented.  The LowE-enhanced image likely represents 
clumpy CSM emission analogous to optical QSF emission.  The 
continuum-dominated image possibly represents emission from diffuse CSM 
swept up by the forward shock.  
\label{tomos}}
\end{figure}

Another likely CSM component, a diffuse component, was discovered when the 
``first light'' observation of \emph{Chandra} revealed a thin, 
continuum-dominated ring of emission around the outer edge of the remnant 
which has been identified as the forward shock \citep{gkr01}.  We find 
similar continuum-dominated knots and filaments across the face of the 
remnant and propose that these are also forward shock filaments seen in 
projection.  These ``projected interior'' filaments are associated with 
edges and filaments of radio emission (DeLaney et al., in preparation) just 
as the ``exterior'' forward shock filaments are associated with the edge of 
the radio plateau \citep{gkr01}.  The projected interior and exterior 
filaments also have the same X-ray spectral shape (bottom panel of Figure 
\ref{spectra}) and they are both associated with steep radio spectral index 
features \citep{ar96}.  The projected interior filaments are not evenly 
distributed across the face of the remnant, but rather are concentrated to 
the west where there is significant X-ray absorption \citep{kra96,wbv02}.  
The ``inward'' motions and the apparent motions faster than free expansion 
for the projected filaments are consistent with the forward shock following 
the path of least resistance through the clumpy medium and showing apparent 
chaotic behavior when seen face on.  We may also be observing a more complex 
pattern involving both motions and changes in synchrotron brightness as 
magnetic fields are amplified in shear layers and in turbulent wakes due to 
the forward shock wrapping around CSM clumps \citep{jk93}.  The interior 
continuum-dominated filaments have also been interpreted as nonthermal 
bremsstrahlung emission associated with the ejecta/reverse shock 
\citep{bwv01}.  The \emph{XMM} hard X-ray continuum images show that the 
continuum-dominated emission in the west is significantly brighter than the 
exterior forward shock filaments possibly indicating that the two 
continuum-dominated populations arise through different emission processes.  
If the interior filaments are instead internal to the SNR, they are unlikely 
to arise through X-ray synchrotron emission because the high magnetic field 
strength will cause rapid aging of electrons at these energies \citep{vl03}.

In summary, the high-resolution proper motion measurements, along with the 
spectral classes, reveal that the X-ray emission can be separated into three 
components each with a distinct kinematic signature.  The ejecta component 
consists of both Si- and Fe-dominated emission populations, is associated 
with optical FMK emission, and has been significantly decelerated relative to 
the FMKs.  This greater deceleration of the X-ray ejecta is likely due to the 
lower densities in this material relative to the optical FMKs.  The 
slow-moving LowE-enhanced emission may be analogous to optical QSF emission 
thus representing a clumpy CSM component.  A diffuse CSM component may be 
associated with continuum-dominated emission from material swept up by the 
forward shock.  In this scenario, the chaotic motions of the interior 
continuum-dominated filaments would be explained as the projected behavior of 
the forward shock interacting with an inhomogeneous CSM.   The exterior 
continuum-dominated filaments have the same expansion rate as the X-ray 
ejecta suggesting a dynamic coupling between the forward shock and the ejecta. 

\acknowledgments

We thank Una Hwang for valuable conversations and manuscript comments.  This 
work was supported at the University of Minnesota under Smithsonian grant 
SMITHSONIAN/GO1-2051A/NASA, NASA grants HST-AR-09537.01 and NAG5-10774 and by 
the Minnesota Supercomputer Institute.

\clearpage

\begin{deluxetable}{cccccccccc}
\tabletypesize{\scriptsize}
\tablecaption{Knot and Filament Proper Motions \label{tbl1}}
\tablewidth{0pt}
\tablehead{
\colhead{\#} & 
\colhead{blc$_x$} & 
\colhead{blc$_y$} &
\colhead{trc$_x$} &
\colhead{trc$_y$} & 
\colhead{$\mu_x$} & 
\colhead{$\mu_y$} &
\colhead{$v_r$} &
\colhead{$\delta v_r$} & 
\colhead{class}
\\
\colhead{ } &
\colhead{$\arcsec$} &
\colhead{$\arcsec$} &
\colhead{$\arcsec$} &
\colhead{$\arcsec$} &
\colhead{$\arcsec$ yr$^{-1}$} &
\colhead{$\arcsec$ yr$^{-1}$} &
\colhead{\% yr$^{-1}$} &
\colhead{\% yr$^{-1}$} &
\colhead{ }
}
\startdata
1 & \phs\phn2.0 & $-$105.3 & \phs\phn8.4 & \phn$-$99.9 & \phs0.01 & $-$0.20 &  \phs0.19 & 0.07 & L \\
2 & \phs60.5 & $-$122.5 & \phs67.4 & $-$115.6 & \phs0.05 & $-$0.01 & \phs0.03 & 0.05 &  M \\
3 & \phs25.6 & \phn$-$83.6 & \phs32.0 & \phn$-$77.2 & \phs0.01 & $-$0.03 & \phs0.03 & 0.09 & M \\
4 & $-$32.0 & \phn$-$65.4 & $-$25.1 & \phn$-$58.1 & $-$0.01 & $-$0.10 & \phs0.15 & 0.11 & S \\
5 & \phn$-$7.4 & \phn$-$33.0 & \phn$-$1.0 & \phn$-$27.1 & $-$0.02 & \phs0.05 & $-$0.15 & 0.24 & L \\
\enddata
\tablecomments{The complete version of this table is in the electronic 
edition of the Journal.  The printed edition contains only a sample.  All 
knots have been assigned a constant error of 0\farcs07 yr$^{-1}$.  Col. 
1. Knot or filament designation number.  Cols. 2-5.  Location of bottom left 
corner and top right corner of box defining the region used to measure each 
knot or filament and extract spectra (arcseconds of offset on epoch 2002 
image with respect to the optical expansion center at 
$23^{\mathrm{h}}23^{\mathrm{m}}27\fs77$, 
$+58\degr48\arcmin49\farcs4$ (J2000) \citep{tfv01}).  Cols. 6-7. Proper 
motion in right ascension and declination ($\arcsec$ yr$^{-1}$).  Positive 
right ascension offset or motion is defined to be westward.  Col. 8. Radial 
component of motion (\% yr$^{-1}$).  Col. 9. Error in radial expansion 
(\% yr$^{-1}$).  Col. 10.  Spectral classification:  S=Si-dominated, 
F=Fe-dominated, L=LowE-enhanced, C=continuum-dominated, M=mixture of two or 
more classes.}
\end{deluxetable}

\clearpage

\end{document}